\documentclass[twocolumn]{jpsj2}
\def\dfrac#1#2{{\displaystyle\frac{#1}{#2}}}
\def\H{{\cal H}}
\def\v#1{\mib #1}

\def\ln{{\rm{ln}}}
\def\A{{\rm{A}}}
\def\B{{\rm{B}}}
\def\AB{{\rm AB}}
\def\BA{{\rm BA}}
\def\JA{J_{{\rm A}}}
\def\JB{J_{{\rm B}}}

\def\DeltaA{\Delta_{{\rm A}}}
\def\DeltaB{\Delta_{{\rm B}}}
\def\const{{{\rm const}}}
\def\rab{r_{\AB}}

\def\runauthor{Kazuo {\sc Hida}}

\title
{
Real Space Renormalization Group Study of the $S=1/2$ XXZ Chains with Fibonacci Exchange Modulation}

\author
{Kazuo {\sc Hida}
\footnote{E-mail: hida@phy.saitama-u.ac.jp}}

\inst
{
Department of Physics, Faculty of Science,\\ Saitama University, Saitama, Saitama 338-8570
}

\recdate
{April 26, 2004
}

\abst
{
Ground state properties of the $S=1/2$ antiferromagnetic XXZ chain with Fibonacci exchange modulation are studied using the real space renormalization group method for strong modulation. The quantum dynamical critical behavior with a new universality class is predicted in the isotropic case. Combining our results with the weak coupling renormalization group results by Vidal {\it et al.}, the ground state phase diagram is obtained.} 
\kword
{
XXZ chain, Fibonacci modulation, quasiperiodicity, real space renormalization group, quantum dynamical critical behavior
}
\begin{document}
\sloppy
\maketitle
\section{Introduction}

The magnetism of quasiperiodic systems has been one of the main subject of continual studies since the discovery of quasicrystals\cite{shecht}. These quasiperiodic systems have an intermediate character between the regular and random systems. Among them, the single particle electronic states in the quasiperiodic chains which are the one dimensional counterpart of the quasicrystals have been intensively studied from theoretical viewpoint.\cite{kkt1,ko1,kst1,hk1} Recently, this problem has been attracting renewed interest following the synthesis of magnetic quasicrystals with well-localized magnetic moments\cite{sato1,sato2}. The theoretical studies of the quantum magnetism in one and two dimensional quasiperiodic systems are also started by many authors\cite{jh1,kh1,kh2,kh3,vieira,vidal1,vidal2,acg,arl,wjh,jag}. 

It is well known that the $S=1/2$ XY chain can be mapped onto the free spinless fermion chain. The problem is, however, not trivial on the quasiperiodic lattice. This problem has been extensively studied by Kohmoto and coworkers\cite{kkt1,ko1,kst1,hk1} by means of the exact renormalization group method. It is shown that the ground state of the XY chain with Fibonacci exchange modulation is critical with finite nonuniversal dynamical exponent. This approach was extended to include other types of aperiodicity and anisotropy\cite{jh1}. It is clarified that the criticality of the Fibonacci XY chain emerges from the marginal nature of the Fibonacci and other precious mean aperiodicity. For the relevant aperiodicity, more singular behavior with a divergent dynamical exponent is realized even for the XY chain\cite{jh1}.

The investigation of the $S=1/2$ Fibonacci Heisenberg chains started quite recently. In the fermionic language, the Ising component of the exchange coupling corresponds to the fermion-fermion interaction, leading to the strong correlation effect which is one of the most important subject of recent condensed matter physics. 
In the weak modulation regime, Vidal and coworkers\cite{vidal1,vidal2} have shown that the Fibonacci modulation is relevant on the basis of the weak coupling renormalization group (WCRG) calculation. The present author carried out the DMRG calculation and investigated the scaling properties of the low energy spectrum\cite{kh1,kh2}.

In the present work, we investigate the $S=1/2$ Fibonacci XXZ chains using the real space renormalization group (RSRG) method\cite{df1} which is valid for the strong exchange modulation. Combining the RSRG and WCRG results, the schematic ground state phase diagram is obtained. Some of the results for the isotropic case are briefly reported in ref. \citen{kh3}. 

This paper is organized as follows. The model Hamiltonian is defined in the next section. The real space renormalization group formulation is given in section 3. In section 4, the solution for the recursion equation is obtained. A new universality class is predicted for the isotropic case. The ground state phase diagram is presented. The last section is devoted to summary and discussion.

\section{Model Hamiltonian}
Our Hamiltonian is given by,
\begin{eqnarray}
\label{ham}
\H &=& \sum_{i=1}^{N-1} J_{\alpha_i}\left[ S^x_{i}S^x_{i+1}+S^y_{i}S^y_{i+1} + \Delta_{\alpha_i} S^z_{i}S^z_{i+1} \right]\\
&& \ \ (J_{\alpha_i} > 0,\ \alpha_i=\A \ \mbox{or} \ \B), \nonumber
\end{eqnarray}
where $\v{S}_{i}$ are the spin 1/2 operators. The exchange couplings $J_{\alpha_i}$ ($=\JA$ or $\JB$) and anisotropy parameters $\Delta_{\alpha_i}$ ($=\DeltaA$ or $\DeltaB$) follow the Fibonacci sequence generated by the substitution rule,
\begin{equation}
\A \rightarrow \A\B, \ \B \rightarrow \A.
\label{subs}
\end{equation}

In the following, we consider the initial Hamiltonian with $\DeltaA=\DeltaB=\Delta$. However, the renormalized values of $\DeltaA$ and $\DeltaB$ are not equal to each other. Therefore we keep the $\alpha_i$-dependence of  $\Delta_{\alpha_i}$ in eq. (\ref{ham}).

\section{Real Space Renormalization Group}
If one of the couplings $\JA$ or $\JB$ is much larger than the other, we can decimate the spins coupled via the stronger exchange coupling and calculate the effective interaction between the remaining spins by the perturbation method with respect to the weaker coupling.\cite{df1} In the following subsections, the cases $\JA << \JB$ and $\JA >> \JB$ are discussed separately.

\begin{figure}
\centerline{\includegraphics[height=70mm]{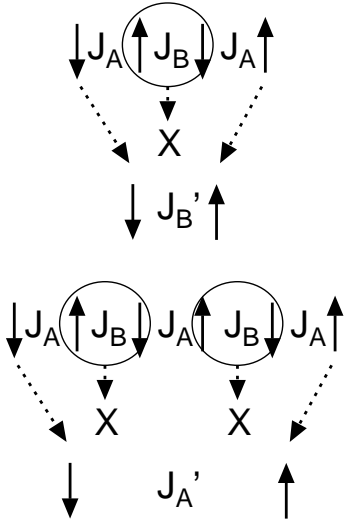}}
\caption{The decimation procedure for $\JA << \JB$.}
\label{deci1}
\end{figure}
\begin{figure}
\centerline{\includegraphics[width=80mm]{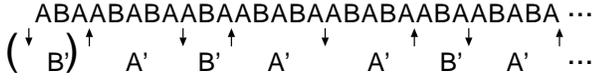}}
\caption{The real space renormalization scheme of the Fibonacci XXZ chain. The letters A and B correspond to the bonds and the up and down arrows to the spins which survive decimation. For $\JA >> \JB$ the end spin and bond in the parenthesis do not appear.}
\label{renorm}
\end{figure}
\subsection{Case $\JA << \JB$}
In this case, the $\JB$ bonds are decimated as shown in Fig. \ref{deci1}. The spin-1/2 degrees of freedom survive on the sites in the middle of the sequence AA. The two kinds of sequences of bonds ABA and ABABA are allowed between two alive spins. There exists one singlet pair in the former sequence while two singlet pairs exist in the latter sequence. Therefore the effective coupling is weaker for the latter case. This decimation process replaces the sequence ABABA by A' and ABA sandwiched by two As by B' resulting in the sequence B'A'B'A'A'B'A'B'A' .. as shown Fig. \ref{renorm}. Except for the B' at the leftmost position, this again gives the Fibonacci sequence as proven in Appendix. The effective coupling can be calculated by the straightforward perturbation theory as,
\begin{eqnarray}
\JA'&=&\frac{\JA^3}{\JB^2(1+\DeltaB)^2} \nonumber\\
\JB'&=&\frac{\JA^2}{\JB(1+\DeltaB)} \nonumber\\
\DeltaA'&=&\frac{\DeltaA^3(1+\DeltaB)^2}{4} \nonumber\\ 
\DeltaB'&=&\frac{\DeltaA^2(1+\DeltaB)}{2} 
\label{largeb}
\end{eqnarray}
up to the lowest order in $\JA$. 

\subsection{Case $\JA >> \JB$}
\begin{figure}
\centerline{\includegraphics[height=70mm]{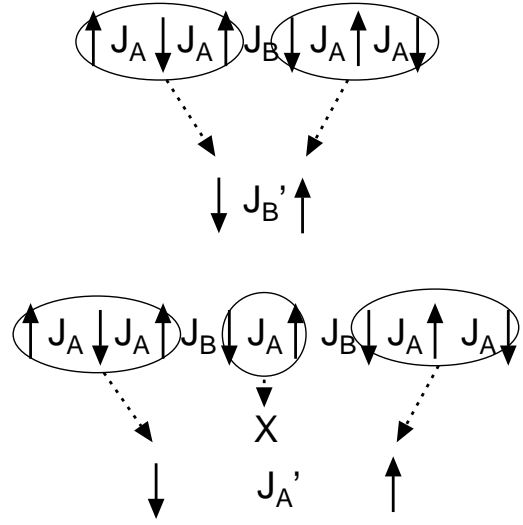}}
\caption{The decimation procedure for $\JA >> \JB$.}
\label{deci2}
\end{figure}
In this case, the $\JA$ bonds are decimated as shown in Fig. \ref{deci2}. The three spins connected by the successive A bonds form a doublet which can be described as a single spin with magnitude 1/2. Therefore the same replacement as the preceding subsection works as well. In this case, no B' appears at the leftmost position as shown in Fig. \ref{renorm}. Again, two kinds of sequences of bonds are allowed between two alive spins, namely B and BAB. There is no singlet pair in the former sequence while one singlet pair exists in the latter sequence. The effective coupling is calculated by the perturbation theory as,
\begin{eqnarray}
\JA'&=&\frac{4\JB^2}{\JA(1+\DeltaA)(8+\DeltaA^2)} \nonumber\\
\JB'&=&\frac{4\JB}{8+\DeltaA^2} \nonumber\\
\DeltaA'&=&\frac{\DeltaB^2\left(\DeltaA+\sqrt{8+\Delta_A^2}\right)^2(1+\DeltaA)}{32} \nonumber\\
\DeltaB'&=&\frac{\DeltaB\left(\DeltaA+\sqrt{8+\Delta_A^2}\right)^2}{16} 
\label{largea}
\end{eqnarray}
up to the lowest order in $\JB$. In this case, the ratio of two couplings is reversed after the first decimation as,
\begin{eqnarray}
\frac{\JA'}{\JB'}&=&\frac{\JB}{\JA(1+\DeltaA)} << 1.
\end{eqnarray}
Therefore, after the second decimation, the formulae in (\ref{largeb}) apply. 

We thus have in general,
\begin{eqnarray}
\JA^{(n+1)}&=&\frac{\JA^{(n)3}}{\JB^{(n)2}\left(1+{\DeltaB}^{(n)}\right)^2}, \label{jarec} \\
\JB^{(n+1)}&=&\frac{\JA^{(n)2}}{\JB^{(n)}\left(1+\DeltaB^{(n)}\right)},\label{jbrec} \\
\DeltaA^{(n+1)}&=&\frac{\DeltaA^{(n)3}\left(1+\DeltaB^{(n)}\right)^2}{4}, \label{anisa}\\
\DeltaB^{(n+1)}&=&\frac{\DeltaA^{(n)2}\left(1+\DeltaB^{(n)}\right)}{2},
\label{anisb} 
\end{eqnarray}
with
\begin{eqnarray}
\JA^{(1)}&=&\frac{\JA^{3}}{\JB^{2}(1+{\DeltaB}^{})^2}, \nonumber\\
\JB^{(1)}&=&\frac{\JA^{2}}{\JB^{}(1+\DeltaB^{})}, \nonumber\\
\DeltaA^{(1)}&=&\frac{\DeltaA^{3}(1+\DeltaB^{})^2}{4}, \nonumber\\
\DeltaB^{(1)}&=&\frac{\DeltaA^{2}(1+\DeltaB^{})}{2}\nonumber
\end{eqnarray}
for $\JB >> \JA$ and 
\begin{eqnarray}
\JA^{(1)}&=&\frac{4\JB^2}{\JA(1+\DeltaA)(8+\DeltaA^2)}, \nonumber\\
\JB^{(1)}&=&\frac{4\JB}{8+\DeltaA^2}, \nonumber\\
\DeltaA^{(1)}&=&\frac{\DeltaB^2\left(\DeltaA+\sqrt{8+\Delta_A^2}\right)^2(1+\DeltaA)}{32}, \nonumber\\
\DeltaB^{(1)}&=&\frac{\DeltaB\left(\DeltaA+\sqrt{8+\Delta_A^2}\right)^2}{16}\nonumber 
\end{eqnarray}
for $\JA >> \JB$ where the variables with $^{(n)}$ refer to the values after $n$-th decimation.

\section{Solutions of the Recursion Equations}
\subsection{XY and isotropic case}

In the XY ($\DeltaA=\DeltaB=0$) and isotropic ($\DeltaA=\DeltaB=1$) case, the recursion equations (\ref{anisa}) and (\ref{anisb}) imply that $\DeltaA$ and $\DeltaB$ are not renormalized. Therefore the recursion equations are simplified as, 
\begin{eqnarray}
\JA^{(n+1)}&=&\frac{\JA^{(n)3}}{\JB^{(n)2}}, \ \ \JB^{(n+1)}=\frac{\JA^{(n)2}}{\JB^{(n)}} ,\ \ (n \geq 1)
\end{eqnarray}
with
\begin{eqnarray}
\JA^{(1)}&=&\dfrac{\JA^{3}}{\JB^{2}},\ \ \JB^{(1)}=\dfrac{\JA^2}{\JB} 
\ \ \mbox{for}\ \ \JA << \JB \nonumber \\
\JA^{(1)}&=&\dfrac{\JB^{2}}{2\JA^{}},\ \ \JB^{(1)}=\dfrac{\JB^{}}{2}\ \ \mbox{for}\ \ \JA >> \JB \nonumber 
\end{eqnarray}
for the XY case and

\begin{eqnarray}
\JA^{(n+1)}&=&\frac{\JA^{(n)3}}{4\JB^{(n)2}}, \ \ \JB^{(n+1)}=\frac{\JA^{(n)2}}{2\JB^{(n)}} \ \ (n \geq 1)
\end{eqnarray}
with
\begin{eqnarray}
\JA^{(1)}&=&\frac{\JA^{3}}{4\JB^{2}}, \ \ \JB^{(1)}=\frac{\JA^{2}}{2\JB} \ \ \mbox{for}\ \ \JA << \JB \nonumber \\
\JA^{(1)}&=&\frac{2\JB^2}{9\JA}, \ \ \JB^{(1)}=\frac{4\JB}{9}\ \ \mbox{for}\ \ \JA >> \JB \nonumber 
\end{eqnarray}
for the isotropic case, 

These recursion equations can be cast into the linear recursion equations for $\ln \JA^{(n)}$ and $\ln \JB^{(n)}$ as follows,
\begin{eqnarray}
\v{X^{(n+1)}}&=&M \v{X^{(n)}}-\v{\alpha} \ \ \mbox{for}\ n \geq 1,
\label{recur_sc}
\end{eqnarray}
where 
\begin{eqnarray}
\v{X^{(n)}}&=&\left(
\begin{array}{l}
\ln \JA^{(n)} \\
\ln \JB^{(n)}
\end{array}\right)\nonumber\\
M&=&\left(
\begin{array}{ll}
3 & -2 \\
2 & -1 
\end{array}
\right) .\nonumber
\end{eqnarray}
The constant vector $\v{\alpha}=^t(\alpha_{\rm A}, \alpha_{\rm B})$ is given by
\begin{eqnarray}
\v{\alpha}&=&\left(
\begin{array}{l}
0 \\
0 
\end{array}
\right) , 
\end{eqnarray}
in the XY case and 
\begin{eqnarray}
\v{\alpha}&=&\left(
\begin{array}{l}
2 \\
1 
\end{array}
\right)\ln 2 , 
\end{eqnarray}
in the isotropic case. 

This equation can be solved easily by noting that
\begin{eqnarray}
M&=&I+O \\
O&=&\left(
\begin{array}{ll}
2 & -2 \\
2 & -2 
\end{array}
\right) ,
\end{eqnarray}
and $I$ is the identity matrix. Because $O^2=0$, the following identity holds
\begin{equation}
M^n = (I + O)^n = I + nO.
\label{zeropower}
\end{equation} 
The $n$-times iteration of eq. (\ref{recur_sc}) gives
\begin{eqnarray}
\lefteqn{\v{X^{(n)}}=M^{n-1}\v{X^{(1)}}-\sum_{k=0}^{n-2}M^k\v{\alpha}}\nonumber\\
&=&(I+(n-1)O)\v{X^{(1)}}-\sum_{k=0}^{n-2}(I+kO)\v{\alpha}\nonumber\\
&=&\left(I+(n-1)O\right)\v{X^{(1)}}\nonumber\\
&-&\left((n-1)I+\frac{(n-2)(n-1)}{2}O\right)\v{\alpha}
\label{recur_sc_sol}
\end{eqnarray}
Namely, we have
\begin{eqnarray}
\lefteqn{
\left(
\begin{array}{l}
\ln \JA^{(n)} \\
\ln \JB^{(n)}
\end{array}\right)
=\left(
\begin{array}{ll}
2n-1 & -2n+2 \\
2n-2 & -2n+3 
\end{array}
\right)
\left(
\begin{array}{l}
\ln \JA^{(1)} \\
\ln \JB^{(1)}
\end{array}\right)}\nonumber\\
&-&
\left(
\begin{array}{ll}
(n-1)^2 & -n^2+3n-2 \\
n^2-3n+2 & -n^2+4n-3 
\end{array}
\right)
\left(\begin{array}{l}
\alpha_{\rm A} \\
\alpha_{\rm B}
\end{array}\right),\nonumber\\
\label{sol}
\end{eqnarray}
for $n \geq 1$. In the following, we discuss the XY and isotropic cases separately.
\subsubsection{XY case}

In the XY case, $\v{\alpha}=0$. Therefore, the last term of (\ref{sol}) vanishes and the solution is given by, 
\begin{eqnarray}
\left(
\begin{array}{l}
\ln \JA^{(n)} \\
\ln \JB^{(n)}
\end{array}\right)
&=&\left(
\begin{array}{ll}
2n-1 & -2n+2 \\
2n-2 & -2n+3 
\end{array}
\right)
\left(
\begin{array}{l}
\ln \JA^{(1)} \\
\ln \JB^{(1)}
\end{array}\right).\nonumber\\
\label{solxy}
\end{eqnarray}
Namely,
\begin{eqnarray}
\dfrac{\JA^{(n)}}{\JA^{(1)}} &=&\left(\dfrac{\JA^{(1)}}{\JB^{(1)}}\right)^{2(n-1)},\\
\dfrac{\JB^{(n)}}{\JB^{(1)}} &=&\left(\dfrac{\JA^{(1)}}{\JB^{(1)}}\right)^{2(n-1)}. 
\end{eqnarray}
For $\JB >> \JA$, this reduces to,
\begin{eqnarray}
\JA^{(n)} &=&\JA\left(\dfrac{\JA}{\JB}\right)^{2n},\\
\JB^{(n)} &=&\JB\left(\dfrac{\JA}{\JB}\right)^{2n}, 
\end{eqnarray}
and for $\JB << \JA$, 
\begin{eqnarray}
\JA^{(n)} &=&\dfrac{\JB^{2}}{2\JA}\left(\dfrac{\JB}{\JA}\right)^{2(n-1)},\\
\JB^{(n)} &=&\dfrac{\JB}{2}\left(\dfrac{\JB}{\JA}\right)^{2(n-1)}. 
\end{eqnarray}

It follows that the ratio $\JA^{(n)}/\JB^{(n)}$ is invariant under renormalization after second step of renormalization as,
\begin{eqnarray}
\frac{\JA^{(n+1)}}{\JB^{(n+1)}}&=&\frac{\JA^{(n)}}{\JB^{(n)}}= ... \nonumber\\
&=&\frac{\JA^{(1)}}{\JB^{(1)}}=\rab^{-1}.
\end{eqnarray}
where $r_{\AB}=\mbox{max}\left(\frac{\JA}{\JB},\frac{\JB}{\JA}\right)$. Therefore, the aperiodicity is marginal for the XY case. This is consistent with the result of the exact renormalization group method\cite{kkt1,ko1,kst1,hk1} and weak coupling renormalization group method.\cite{vidal1,vidal2} This also implies that the perturbation approximation remains valid if it is in the first step of renormalization.

As explained in Appendix, our decimation rule 
\begin{displaymath}
\A\B\A\B\A \rightarrow \A, \ \ \A\B\A \rightarrow \B 
\label{dec3}
\end{displaymath}
is almost equivalent to the usual 3-step deflation process
\begin{displaymath}
\A\B\A\A\B \rightarrow \A, \ \ \A\B\A \rightarrow \B 
\label{sub3}
\end{displaymath}
except for a single B which remains at the left end. It should be noted that the length of the $3n$-th Fibonacci sequence is equal to the Fibonacci number $F_{3n}$ which grows as $\phi^{3n}$ for large $n$ where $\phi$ is the golden mean ($=\frac{1+\sqrt{5}}{2}$). This implies that the finite Fibonacci sequence of length $N$ reduces to a single spin pair coupled via $\JB^{(n)}$ after $n= \ln N / (3\ln \phi)$  steps of decimations. Therefore, the smallest energy scale $\Delta E$ for the finite Fibonacci XY chain with length $N$ scales as,
\begin{eqnarray}
\Delta E &\sim &N^{-z} \ \ \mbox{with} \ \ z=\dfrac{2}{3\ln \phi}\ln r_{\AB}.
\label{dynex} 
\end{eqnarray}
This reproduces the result of Kohmoto and coworkers\cite{kkt1,ko1,kst1,hk1} in the limit $r_{\AB} >> 1$.

\subsubsection{Isotropic case}
In the isotropic case, $\v{\alpha}=^t(2, 1)\ln2$ and the solution is given by, 
\begin{eqnarray}
\lefteqn{\left(
\begin{array}{l}
\ln \JA^{(n)} \\
\ln \JB^{(n)}
\end{array}\right)
=\left(
\begin{array}{ll}
2n-1 & -2n+2 \\
2n-2 & -2n+3 
\end{array}
\right)
\left(
\begin{array}{l}
\ln \JA^{(1)} \\
\ln \JB^{(1)}
\end{array}\right)\nonumber}\\
&-&
\left(
\begin{array}{ll}
(n-1)^2 & -n^2+3n-2 \\
n^2-3n+2 & -n^2+4n-3 
\end{array}
\right)
\left(\begin{array}{l}
2 \\
1
\end{array}\right)\ln2,\nonumber\\
&&
\label{soliso}
\end{eqnarray}
It should be noted that the last term is most dominant in the limit $n \rightarrow \infty$. We thus have, 
\begin{eqnarray}
\dfrac{\JA^{(n)}}{\JA^{(1)}} &=&\left(\dfrac{\JA^{(1)}}{\JB^{(1)}}\right)^{2(n-1)}2^{-n(n-1)}, \\
\dfrac{\JB^{(n)}}{\JB^{(1)}} &=&\left(\dfrac{\JA^{(1)}}{\JB^{(1)}}\right)^{2(n-1)}2^{-(n-1)^2}. 
\end{eqnarray}
For $\JB >> \JA$, this reduces to,
\begin{eqnarray}
\JA^{(n)} &=&\JA\left(\dfrac{\JA}{\JB}\right)^{2n}2^{-n(n+1)},\\
\JB^{(n)} &=&\JB\left(\dfrac{\JA}{\JB}\right)^{2n}2^{-n^2}, 
\end{eqnarray}
and for $\JB << \JA$, 
\begin{eqnarray}
\JA^{(n)} &=&\dfrac{8\JB}{9}\left(\dfrac{\JB}{\JA}\right)^{2n-1}2^{-n(n+1)},\\
\JB^{(n)} &=&\dfrac{8\JA}{9}\left(\dfrac{\JB}{\JA}\right)^{2n-1}2^{-n^2} .
\end{eqnarray}

It follows that the ratio $\JA^{(n)}/\JB^{(n)}$ decreases under renormalization as,
\begin{eqnarray}
\frac{\JA^{(n+1)}}{\JB^{(n+1)}}&=&\frac{1}{2}\frac{\JA^{(n)}}{\JB^{(n)}} \ \ (n \geq 1)
\end{eqnarray}
and for the first step, 
\begin{eqnarray}
\frac{\JA^{(1)}}{\JB^{(1)}}&=&\frac{1}{2}\rab^{-1}.
\end{eqnarray}
Therefore the aperiodicity is relevant in the isotropic case. This is consistent with the result of the WCRG. This also implies that the perturbation approximation becomes even more accurate as the renormalization proceeds. 

Taking into account that $n = \ln N /(3\ln \phi)$ the smallest energy scale $\Delta E$ for the finite Fibonacci chain with length $N$ scales as,
\begin{eqnarray}
\Delta E &\sim &2^{-n^2}\sim \exp\left({-(\ln N /3\ln \phi)^2\ln 2}\right)\nonumber\\
&=& N^{-\kappa\ln N }\ \ \mbox{with}\ \ \kappa\equiv \ln 2/(3\ln \phi)^2
\label{true}
\end{eqnarray}
for large enough $N$, irrespective of the value of $\rab$. This type of quantum dynamical critical behavior, which implies the logarithmic divergence of dynamical exponent, is not yet reported in any regular, quasiperiodic and random systems in the field of one-dimensional quantum magnetism.

This size dependence of the energy scale implies that the number of the magnetic excited states $ND(\Delta E)d\Delta E$ in the energy range $\Delta E \sim \Delta E + d\Delta E$ can be written in the scaling form in terms of the scaled variable $x=N\exp\left(-\sqrt{\frac{1}{\kappa}\ln \frac{1}{\Delta E}}\right)$ as,
\begin{eqnarray}
\lefteqn{ND(\Delta E)d\Delta E \sim f\left(x\right)dx} \nonumber\\
&\sim& Nf\left(N\exp\left(-\sqrt{\frac{1}{\kappa}\ln \frac{1}{\Delta E}}\right)\right)\nonumber\\
&\times&\frac{\exp\left(-\sqrt{\frac{1}{\kappa}\ln \frac{1}{\Delta E}}\right)}{2\kappa\Delta E \sqrt{\frac{1}{\kappa}\ln \frac{1}{\Delta E}}}d\Delta E 
\end{eqnarray}
with a scaling function $f(x)$. Because the density of state per site $D(\Delta E)$ should be finite in the thermodynamic limit $N \rightarrow \infty$, the scaling function $f(x)$ should tend to a finite value as $x \rightarrow \infty$ ($N \rightarrow \infty$). Therefore we find,
\begin{eqnarray}
D(\Delta E)d\Delta E &\sim& \frac{\exp\left(-\sqrt{\frac{1}{\kappa}\ln \frac{1}{\Delta E}}\right)}{2\kappa\Delta E \sqrt{\frac{1}{\kappa}\ln \frac{1}{\Delta E}}}d\Delta E 
\end{eqnarray}
for large enough $N$. Accordingly, the low temperature magnetic specific heat $C(T)$ at temperature $T$ should behave as,
\begin{eqnarray}
C(T) &\sim& \frac{\partial}{\partial T}N\int_0^T \Delta E D(\Delta E) d\Delta E \sim NTD(T) \nonumber\\ 
&\sim& \frac{N}{2\kappa \sqrt{\frac{1}{\kappa}\ln \frac{1}{T}}}\exp\left(-\sqrt{\frac{1}{\kappa}\ln \frac{1}{T}}\right).
\label{spe}
\end{eqnarray}
The magnetic susceptibility $\chi(T)$ at temperature $T$ should be the Curie contribution from the spins alive at the energy scale $T$. The number $n_s(T)$ of such spins is given by 
\begin{eqnarray}
n_s(T)&\sim& 2N\int_0^T D(\Delta E) d\Delta E, 
\end{eqnarray}
because two spins are excited by breaking a single effective bond with effective exchange energy less than $T$. Therefore the low temperature magnetic susceptibility $\chi$ behaves as,
\begin{eqnarray}
\chi(T)&\sim& \frac{2N}{4T}\int_0^T \frac{\exp\left(-\sqrt{\frac{1}{\kappa}\ln \frac{1}{\Delta E}}\right)}{2\kappa\Delta E \sqrt{\frac{1}{\kappa}\ln \frac{1}{\Delta E}}}d\Delta E \nonumber \\ 
&\sim& \frac{N\exp\left(-\sqrt{\frac{1}{\kappa}\ln \frac{1}{T}}\right)}{2T}.
\label{sus}
\end{eqnarray}

In ref. \citen{kh1}, the present author carried out the DMRG calculation for the Fibonacci antiferromagnetic Heisenberg chains. We performed the finite size scaling analysis of the lowest energy gap $\Delta E$ based on the {\it assumption} that it will behave in the same way as the XY chain with relevant aperiodicity, namely as $\Delta E \sim \exp(-cN^{\omega})$. However, the present analysis suggests the different behavior (\ref{true}). Therefore we tried to replot the previous data in ref. \citen{kh1} using the scaling (\ref{true}). Unfortunately, however, the fit is very poor. The reason will be understood in the following way.

In the present RSRG analysis, we started the RG transformation based on the perturbation theory assuming that one of the exchange coupling is much larger than the other ($\rab >> 1$). However, the DMRG calculation in ref. \citen{kh1} is carried out for relatively small $\rab$. Therefore in the early stage of renormalization, the strong coupling recursion formulae are not valid. To supplement this point, we introduce the phenomenological modified recursion formulae
\begin{eqnarray}
\label{mja}\JA^{(n+1)}&=&\frac{\JA^{(n)3}}{\JB^{(n)2}}\exp(-\alpha_{\rm A}), \\
\label{mjb}\JB^{(n+1)}&=&\frac{\JA^{(n)2}}{\JB^{(n)}}\exp(-\alpha_{\rm B}), 
\end{eqnarray}
where $\v{\alpha}$ depends on the ratio $\JA^{(n)}/\JB^{(n)}$. This leads to the modified recursion equation for $\v{X}^{(n)}$ as,
\begin{eqnarray}
\v{X^{(n+1)}}&=&M \v{X^{(n)}}-\v{\alpha(X^{(n)}_{\AB})}
\label{recur_scm}
\end{eqnarray}
where $\v{\alpha}$ is a function of $X^{(n)}_{\AB}\equiv X^{(n)}_A-X^{(n)}_B$. If the function $\v{\alpha}(X)$ is approximated by a linear function of $X$, we have 
\begin{eqnarray}
\v{X^{(n+1)}}&=&M \v{X^{(n)}}-(X^{(n)}_{\AB}-X^{(0)}_{\AB})\v{\gamma}\nonumber\\
&-&\v{\alpha(X^{(0)}_{\AB})} ,
\label{recur_scml}
\end{eqnarray}
which can be rewritten as,
\begin{eqnarray}
\v{X^{(n+1)}}&=&M_{\rm m} \v{X^{(n)}}-\v{\alpha}_{\rm m}
\label{recur_scml2}
\end{eqnarray}
where
\begin{eqnarray}
M_{\rm m}&=&\left(
\begin{array}{ll}
3-\gamma_A & -2+\gamma_A \\
2-\gamma_B & -1+\gamma_B 
\end{array}
\right) \nonumber\\
\v{\alpha}_{\rm m}&=&\v{\alpha(X^{(0)}_{\AB})}-X^{(0)}_{\AB}\v{\gamma}, \nonumber\\
\v{\gamma}&=&(\gamma_{\rm A}, \gamma_{\rm B}). \nonumber
\end{eqnarray}
The iterative solution is given by,
\begin{eqnarray}
\v{X^{(n)}}&=&M_{\rm m} \v{X^{(n-1)}}-\v{\alpha}_{\rm m}\nonumber\\
&=&M_{\rm m}^{n-1}\v{X^{(1)}}-\sum_{k=0}^{n-2}M_{\rm m}^k\v{\alpha}_{\rm m}.
\label{recur_sc_solm}
\end{eqnarray}
In this case, the equation corresponding to (\ref{zeropower}) does not hold for $M_{\rm m}$. By elementary algebra, we find the eigenvalues of $M_{\rm m}$ are $1+\gamma_{\BA}\ (\gamma_{\BA}\equiv \gamma_{\B}-\gamma_{\A})$ and unity. Therefore, if $\gamma_{\B}>\gamma_{\A}$, $\v{X}^{(n)}$ grows as,
\begin{eqnarray}
\v{X^{(n)}}&\sim& \lambda_{\rm m}^n \v{u}_{\rm m} \ \ \mbox{with}\ \ \lambda_{\rm m} \equiv 1+\gamma_{\BA}
\end{eqnarray}
where $\v{u}_{\rm m}$ is the eigenvector corresponding to $\lambda_{\rm m}$. This implies 
\begin{eqnarray}
\ln \JA^{(n)}&\sim& \ln \JB^{(n)}\sim \lambda_{\rm m}^n .
\label{initial}
\end{eqnarray}

\begin{figure}
\centerline{\includegraphics[width=80mm]{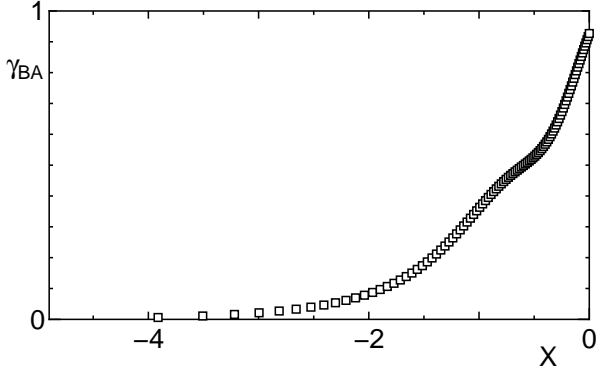}}
\caption{The $X$-dependence of $\gamma_{\BA}$ calculated from finite cluster numerical diagonalization. }
\label{gamma}
\end{figure}

The effective value of $\gamma_{\BA}$, which depends on $X=\ln(\JA/\JB)$, is estimated in the following way. (In the remainder of this subsection, we omit the superscript $^{(n)}$ and distinguish the renormalized values by $'$.) Let us concentrate on the case $\JA < \JB$, because after the second step this inequality always holds. To determine $\alpha_{\A}$ for $\JA < \JB$, we numerically diagonalize the cluster BAABABAAB. This is the largest cluster which reduces to a single A'-bond after decimation. Therefore the effective coupling $\JA'$ of the A'-bond is estimated from the singlet-triplet energy gap of this cluster. The parameter $\alpha_{\rm A}$ is determined through the formula (\ref{mja}). To determine $\alpha_{\B}$, the same procedure is repeated with the cluster BABAABAABAB using the formula (\ref{mjb}). By the numerical differentiation of $\alpha_{\B}-\alpha_{\A}$ with respect to $X$, the effective value of $\gamma_{\BA}$ is obtained as shown in Fig. \ref{gamma}. It is seen from this figure that $\gamma_{\BA} > 0$. Therefore, starting from finite $\JA/\JB$, in the early stage of renormalization, the effective coupling renormalizes as eq. (\ref{initial}) similarly to the case of XY model with relevant aperiodicity.\cite{jh1} In terms of the system size dependence, this implies that the finite size energy gap behaves as, 
\begin{eqnarray}
\label{short}
\ln\Delta E&\sim& \lambda_{\rm m}^{\frac{\ln N}{3\ln \phi}} \sim N^{\frac{\ln \lambda_{\rm m}}{3\ln \phi}}\sim N^{\omega} \\
&& \ \mbox{with} \ \ \omega\equiv\frac{\ln \lambda_{\rm m}}{3\ln \phi}\nonumber
\end{eqnarray}
for relatively small systems. This is the reason why we have observed the behavior (\ref{short}) in DMRG calculation in ref. \citen{kh1}. As the renormalization precess proceeds, the ratio $\JA/\JB$ decreases and $\gamma_{\BA}$ tends to zero. Therefore, the formula (\ref{true}) is recovered in the true asymptotic limit.

\subsection{XXZ case}
\subsubsection{XY-like case $(|\Delta|< 1)$}
For $|\Delta|< 1$, the anisotropy parameters are always renormalized to $\DeltaA=\DeltaB=0$ as is obvious from the recursion relations (\ref{anisa}) and (\ref{anisb}). Therefore the XY type behavior $\Delta E \sim N^{-z}$ with finite dynamical exponent $z$ is always expected. 

\begin{figure}
\centerline{\includegraphics[width=80mm]{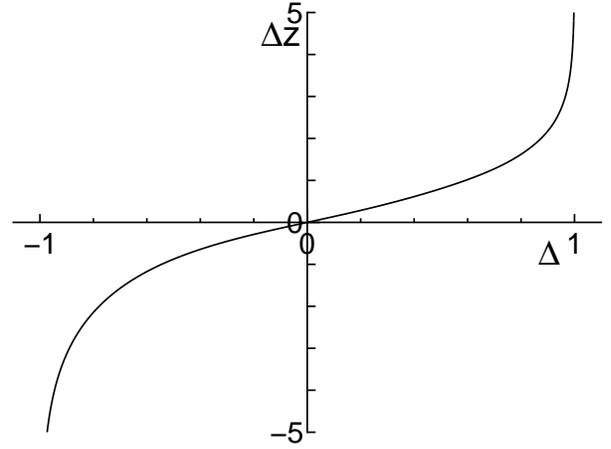}}
\caption{The shift $\Delta z$ of the dynamical exponent $z$ relative to the XY value $z_{\rm XY}$ plotted against $\Delta$. }
\label{deltaz}
\end{figure}

The exponent $z$ is determined by substituting the fixed point values of $\JA/\JB$ into (\ref{dynex}) as, 
\begin{eqnarray}
z&=&\dfrac{2}{3\ln \phi}\ln \dfrac{\JB^{(\infty)}}{\JA^{(\infty)}}
\label{dynexxxz} 
\end{eqnarray}
From (\ref{jarec}) and (\ref{jbrec}), the recursion formula for the ratio $\frac{\JB^{(n)}}{\JA^{(n)}}$ is given by,
\begin{eqnarray}
\dfrac{\JB^{(n)}}{\JA^{(n)}}&=&\dfrac{\JB^{(n-1)}}{\JA^{(n-1)}}{(1+{\DeltaB}^{(n-1)})} 
\end{eqnarray}
whose solution is given by,
\begin{eqnarray}
\dfrac{\JB^{(n)}}{\JA^{(n)}}&=&\dfrac{\JB^{(1)}}{\JA^{(1)}}\displaystyle\prod_{k=1}^{n-1} (1+{\DeltaB}^{(k)})\nonumber\\
&=& r_{\AB}\displaystyle\prod_{k=0}^{n-1} (1+{\DeltaB}^{(k)}) 
\end{eqnarray}
where we have taken into account $\DeltaA^{(0)}=\DeltaB^{(0)}=\Delta$. Therefore the dynamical exponent $z$ is renormalized with respect to the XY value $z_{\rm XY}$ as 
\begin{eqnarray}
z&=&z_{\rm XY} + \Delta z \nonumber \\
z_{\rm XY}&=&\dfrac{2}{3\ln \phi}\ln r_{\AB} \nonumber\\
\Delta z&=&\dfrac{2}{3\ln \phi}\ln \prod_{k=0}^{\infty} (1+{\DeltaB}^{(k)}) 
\end{eqnarray}
The values of $\Delta z$ are calculated for $-1 < \Delta < 1$ as shown in Fig. \ref{deltaz}. For $0 < \Delta < 1$, the dynamical exponent is enhanced and diverges at $\Delta =1$, while it is reduced for $ -1 < \Delta < 0$. It should be also noted that $\Delta z$ does not depend on the ratio $\JA/\JB$.

\begin{figure}
\centerline{\includegraphics[width=80mm]{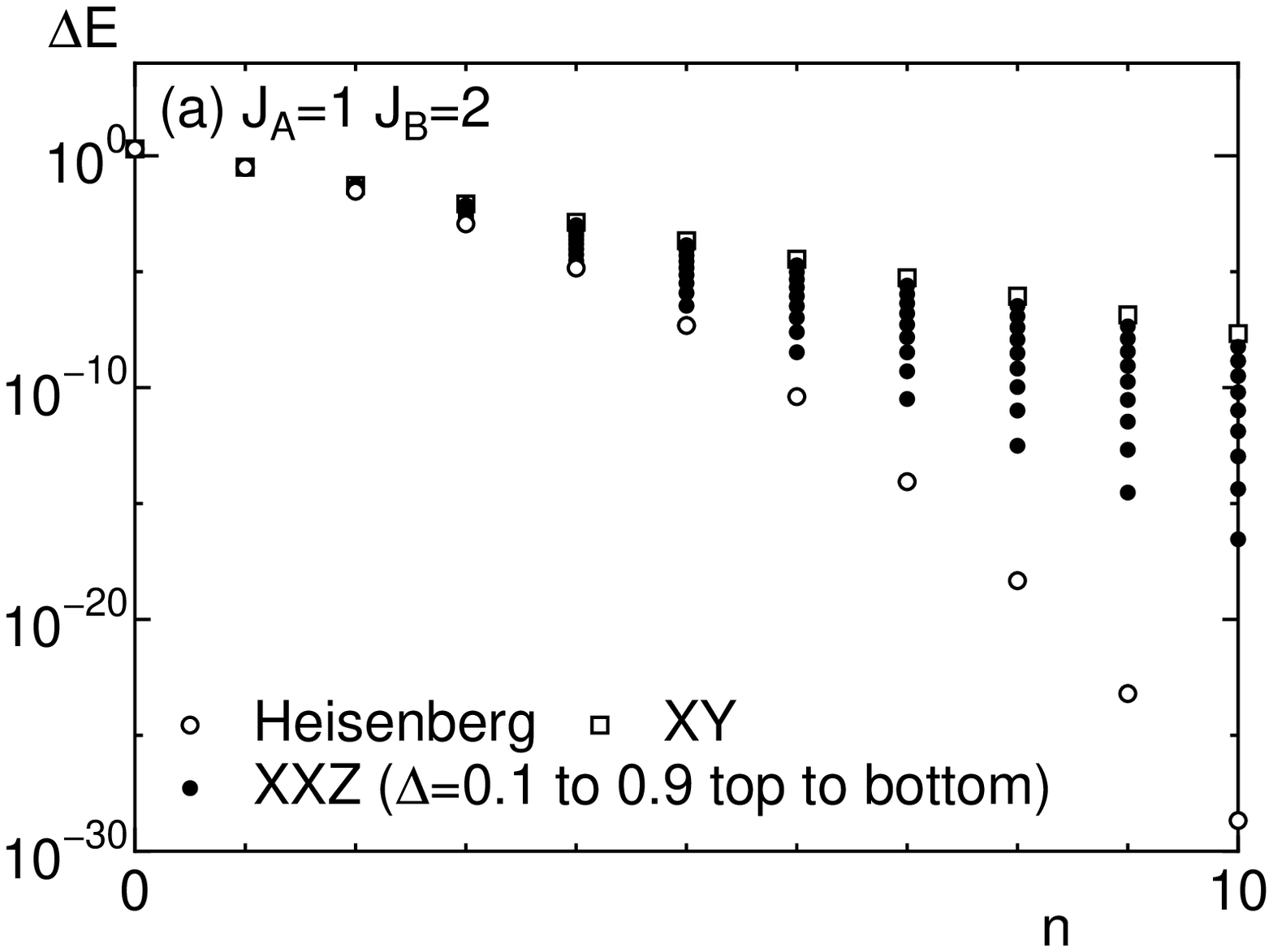}}
\centerline{\includegraphics[width=80mm]{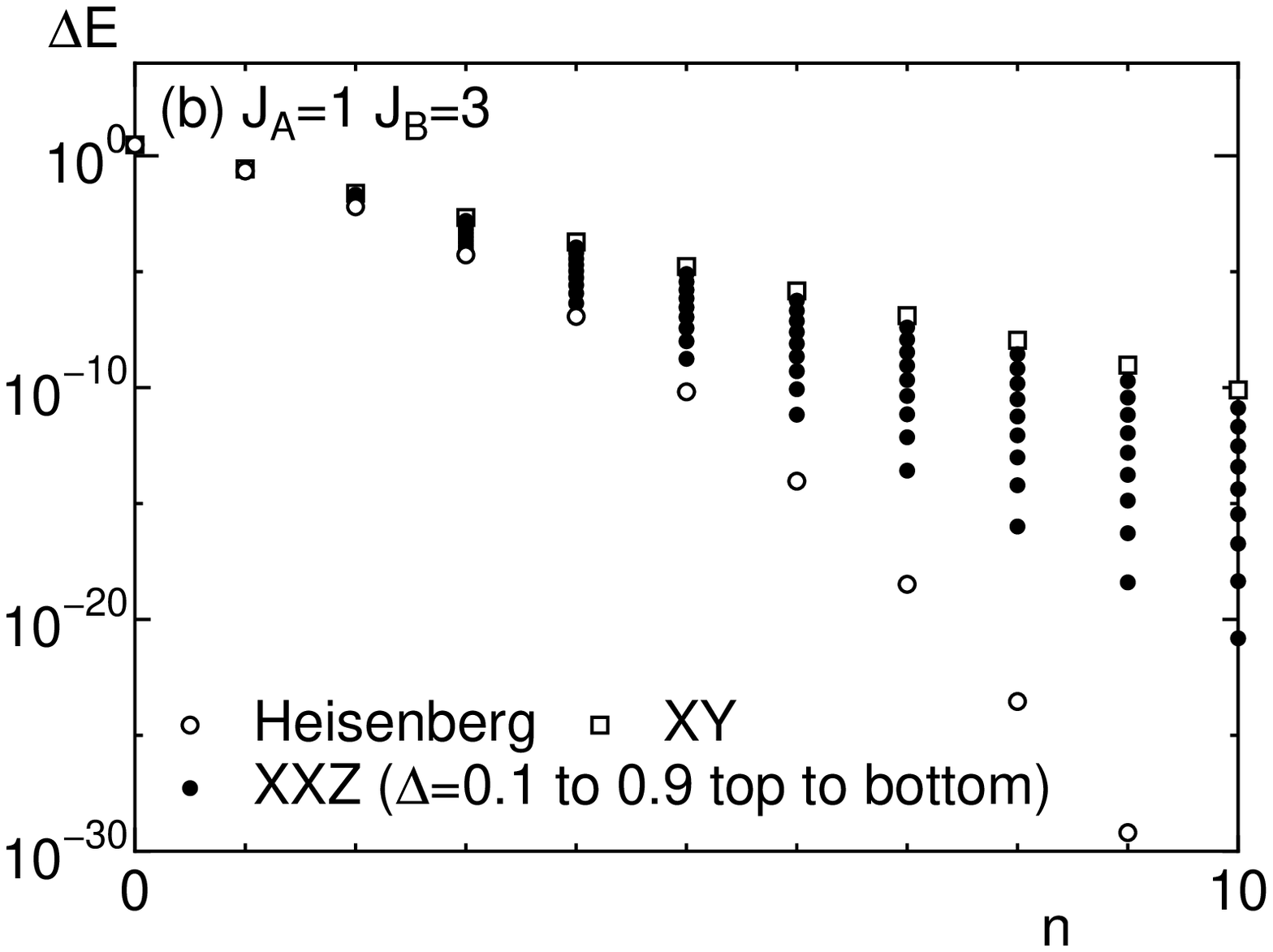}}
\caption{The renormalization step dependence of the energy scale $\Delta E$ obtained by the numerical renormalization group method with (a) $\JB/\JA=2$ and (b) $\JB/\JA=3$. The renormalization step $n$ should be interpreted as $\ln N/(3\ln \phi)$ in terms of the system size $N$.}
\label{numren}
\end{figure}

In the region, $0 < \Delta < 1$, the Fibonacci modulation is always relevant within the WCRG calculation. Therefore, the system parameters are renormalized from the weak modulation regime to the strong modulation regime where our RSRG approach is valid. After enough steps of renormalization, the Fibonacci XXZ model with $0 < \Delta < 1$ is renormalized to the XY chain with Fibonacci modulation and the energy gap scales as $\Delta E \sim N^{-z}$ with nonuniversal dynamical exponent $z$. This looks again in contradiction with the DMRG data in ref. \citen{kh2} which are well fitted by $\ln \Delta E \sim -N^{\omega}$. To clarify this point, we carried out the numerical RSRG calculation using the clusters used in the last section. The effective coupling and anisotropy parameters are determined from the excitation spectrum of the clusters in each step of renormalization. The results are shown in Figs. \ref{numren}(a,b). For $0 < \Delta < 1$, the effective coupling scales as $N^{-z}$ for large enough $N$. However, for small system sizes, the behavior of $\Delta E$ is close to the isotropic case for non-vanishing $\Delta$. This is clearly observed for relatively large $\Delta$ in Figs. \ref{numren}(a,b). This explains why the DMRG data for the energy gap was fitted well by the formula $\ln\Delta E \sim N^{\omega}$.\cite{kh2}

In the region $-1<\Delta< 0$, the situation is more complicated. For large enough $\rab$, the anisotropy $\Delta$ is again renormalized to small values. Therefore, the Fibonacci XY line is approached and the scaling relation $\Delta E \sim N^{-z}$ with nonuniversal dynamical exponent $z$ should hold. 

\begin{figure}
\centerline{\includegraphics[width=70mm]{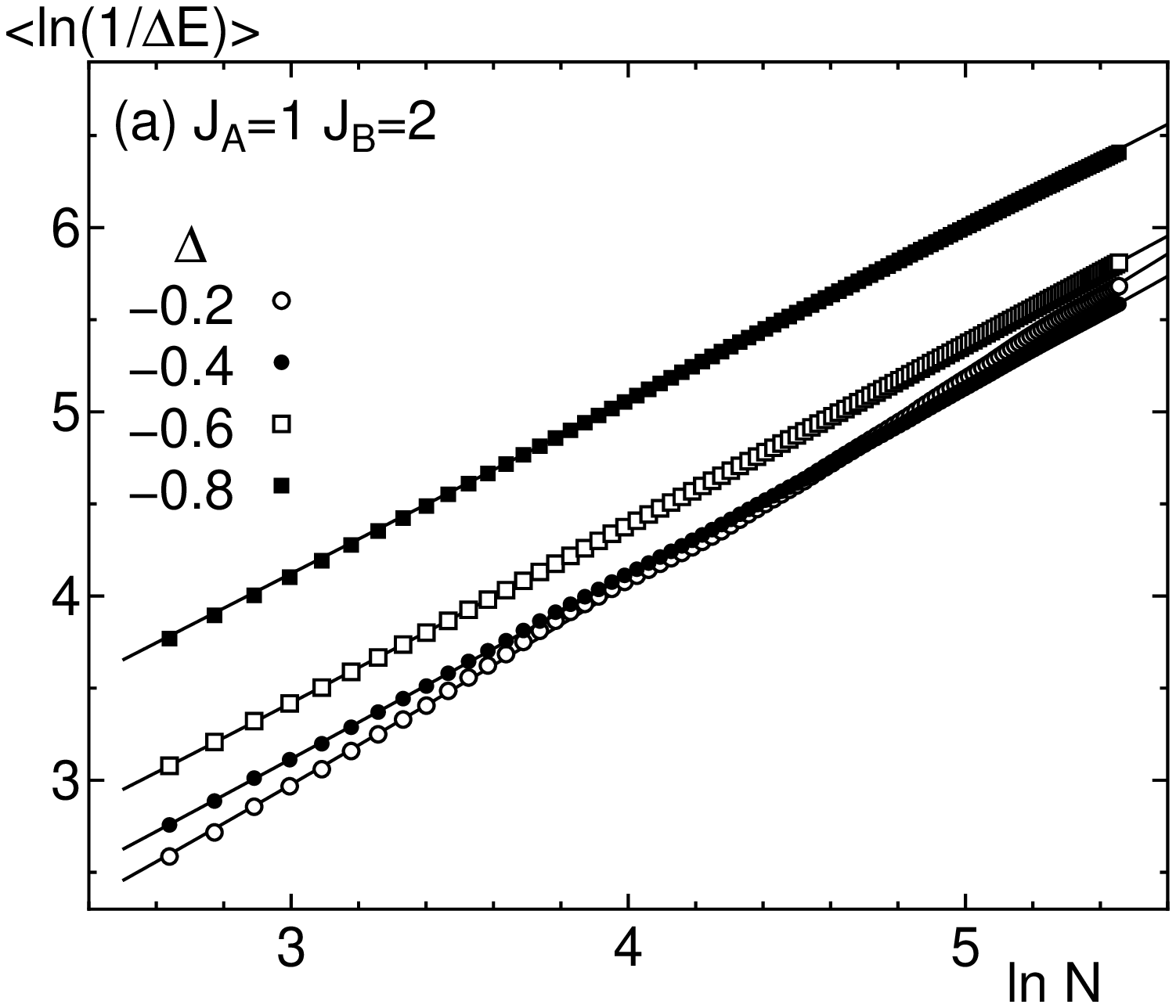}}
\centerline{\includegraphics[width=70mm]{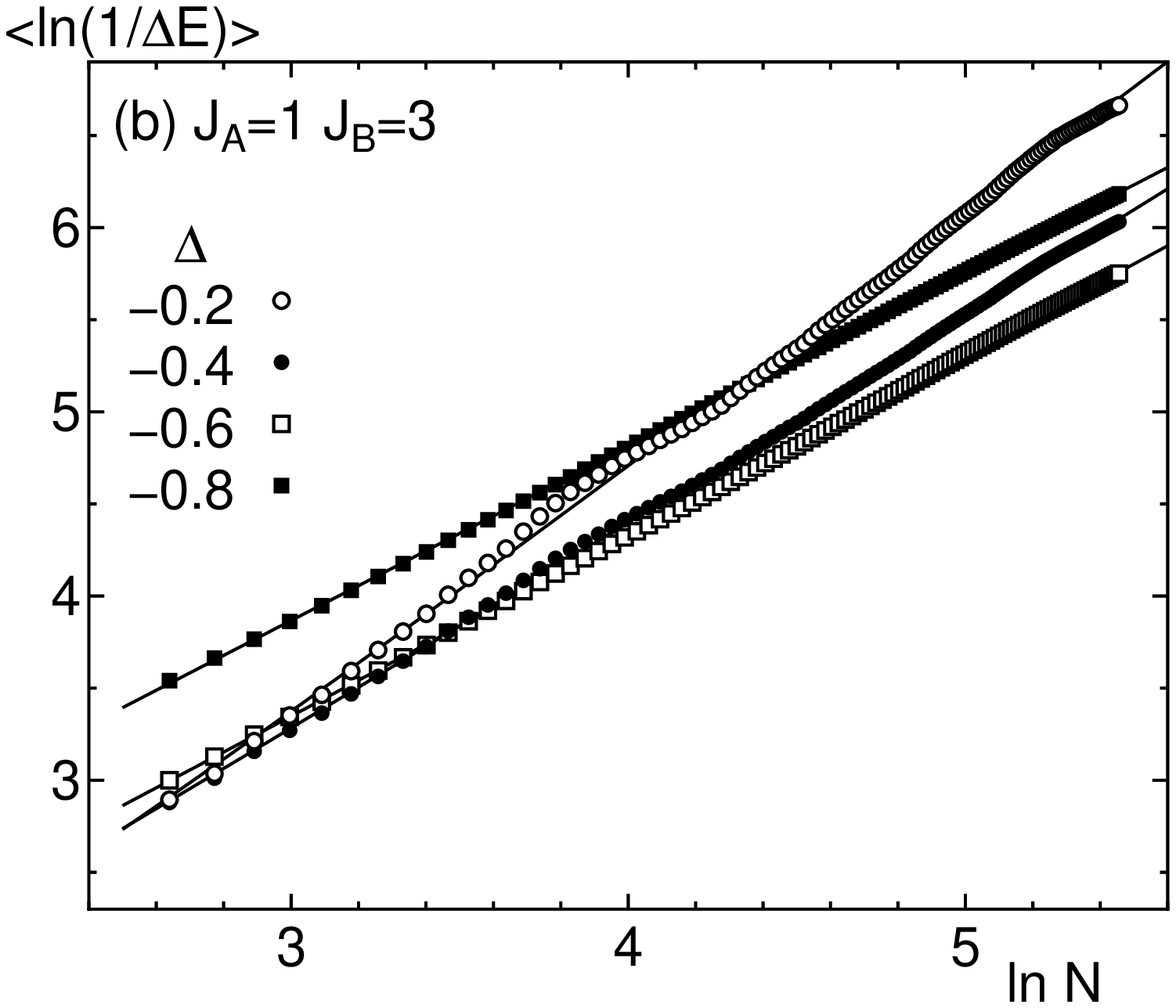}}
\caption{The $N$-dependence of $<\ln(1/\Delta)>$ of the Fibonacci XXZ chain with negative $\Delta$ for (a) $\JA=1$ $\JB=2$ and (b) $\JA=1$ $\JB=3$ by the DMRG method plotted against $\ln N$.}
\label{dmxy}
\end{figure}

According to the WCRG analysis by Vidal and coworkers\cite{vidal1,vidal2}, however, the Fibonacci modulation is irrelevant for $-1 < \Delta < 0$ as far as the modulation amplitude is small. Therefore the ground state should be renormalized to the Luttinger liquid state without aperiodic modulation. This looks inconsistent with our RSRG conclusion. However, as $\Delta$ approaches $-1$, even for large $r_{\AB}$, the perturbation approach breaks down because the energy denominators in (\ref{largeb}) and (\ref{largea}) becomes larger than the strength of perturbation. The limit of reliability of the present strong coupling approximation is given by $r_{\AB} > r_{\rm cr} \simeq (1+\Delta)^{-1}$. For $r_{\AB} < r_{\rm cr}$, the WCRG prediction is more reliable and we expect the Luttinger liquid behavior $\Delta E \sim N^{-1}$, $C \sim T$ and $\chi \sim \const$. For $r_{\AB} > r_{\rm cr}$, the marginal behavior $\Delta E \sim N^{-z}$ , $C \sim T^{1/z}$ and $\chi \sim T^{1/z-1}$ is expected. Actually the DMRG data for $\JB/\JA=2$ and 3 with negative $\Delta$ shown in Fig. \ref{dmxy} confirms the clear deviation from the Luttinger liquid behavior $\Delta E \sim N^{-1}$ for small negative values of $\Delta$. The details of the DMRG calculation are as described in refs. \citen{kh1} and \citen{kh2} except that the maximum number of states retained in each subblock is 160. This value is required due to the large modulation strength.

\subsubsection{Ising-like case : $(|\Delta | > 1)$}

For $\Delta < -1$, it is obvious that the ground state is the ferromagnetic state. For $\Delta > 1$, the anisotropy $\DeltaA$ and $\DeltaB$ are renormalized to infinity as seen from (\ref{anisa}) and (\ref{anisb}). Therefore the ground state is the N\'eel state. However, in this case, the perturbational recursion equation breaks down because $\JA^{(n)}\DeltaA^{(n)}$ becomes larger than $\JB^{(n)}$ for large enough $n$ if our recursion equation is naively solved. This is reasonable because the N\'eel state is not adiabatically connnected to the dimer state which is the starting point of our RSRG scheme. Therefore, it is not possible to make further quantitative predictions in this regime within the RSRG scheme.

\subsubsection{Phase Diagram}

\begin{figure}
\centerline{\includegraphics[width=80mm]{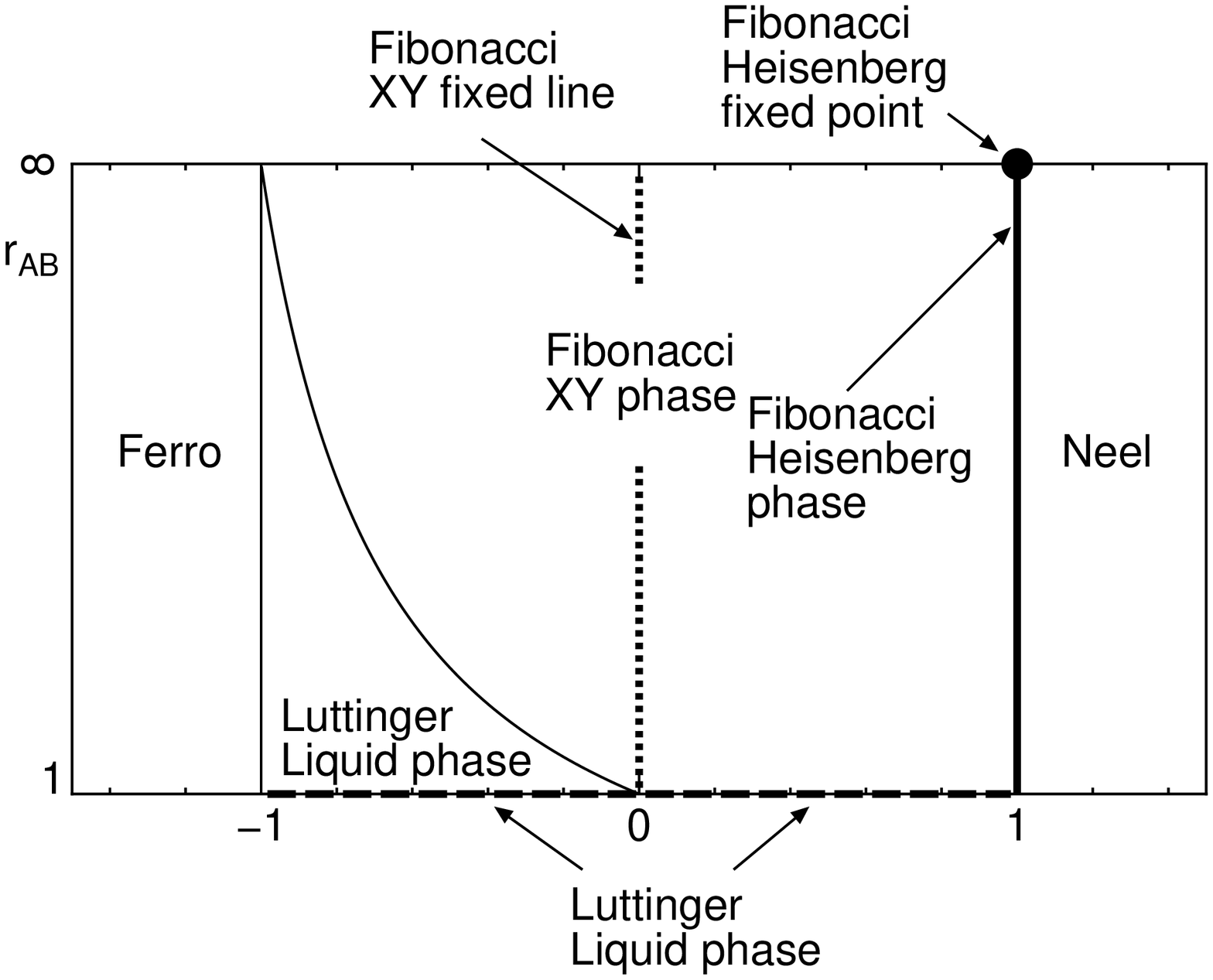}}
\caption{The schematic ground state phase diagram of the Fibonacci XXZ chain. }
\label{phase0}
\end{figure}

All the above arguments are summarized as the schematic ground state phase diagram in Fig. \ref{phase0}. The thin solid curve corresponds to the line $r_{\AB} =r_{\rm cr} \sim (1+\Delta)^{-1}$ which separates the Luttinger liquid phase and the Fibonacci XY phase. The thick solid line is the isotropic Fibonacci Heisenberg phase with scaling $\Delta E \sim N^{-\kappa\ln N }$. The thick broken line belongs to the Luttinger liquid phase. The thick dotted line is the fixed line of the marginal Fibonacci XY state. The N\'eel state and the ferromagnetic state appear for $\Delta > 1$ and $\Delta < -1$, respectively.
\section{Summary and Discussion}

The ground state properties of the $S=1/2$ Fibonacci XXZ chain are investigated by means of the RSRG method. In the isotropic case $\Delta=1$, it is found that the energy gap scales as $N^{-\kappa \ln N}$ which implies the logarithmic divergence of the dynamical exponent. For $-1 < \Delta <1$, the RSRG flow converges to the Fibonacci XY-line. Therefore, the marginal behavior with finite dynamical exponent $z$ is expected as far as the RSRG scheme is reliable. With the help of the WCRG results by Vidal and coworkers\cite{vidal1,vidal2}, which is valid in the weak modulation regime with $-1 < \Delta < 1$, the ground state phase diagram is determined. The previous DMRG results by the present author are reinterpreted from the viewpoint of the present theory and understood as the finite size crossover effect.

We have found a new quantum dynamical critical behavior (\ref{true}) which was so far unknown in the field of quantum many body problem. Similar 'singular dynamic scaling' is, however, known since 80's for the classical Ising model on the percolation clusters with Glauber dynamics.\cite{hen,her,ram} In spite of the geometrical self-similarity common to this classical model and our quantum model, they look very different in many aspects. Although the underlying physics which leads to the similarity in the dynamical critical behavior of both systems is unclear, it would be possible that a kind of hidden symmetry underlies in the problem of the Ising model on the percolation clusters considering that the singular scaling (\ref{true}) in our model only occurs in the isotropic case ($SU(2)$ symmetric case). Further investigation on this point might lead to a more profound understanding of the dynamical critical behavior of both systems.

It is well known that the RSRG method is suitable for the random spin chain problem.\cite{df1,hy1,hy2,wes} The present calculation demonstrates that the RSRG method is also a powerful tool for the investigation of the quasiperiodic chains. Actually, after this work is almost completed, the preprint by Vieira\cite{vieira} appeared in e-print archive in which some of the present results are derived. In addition, Vieira satisfactorily applied this method to the Heisenberg chains with relevant aperiodicity. Similar approach is also applied to the two dimensional quasicrystal.\cite{jag} We thus expect the RSRG method is widely applicable to various problems in the field of quantum magnetism in quasiperiodic systems.

The author would like to thank Dr. C. L. Henley for drawing his attention to refs. \citen{hen,her,ram} and for enlightening comments. The numerical computation in this work has been carried out using the facilities of the Supercomputer Center, Institute for Solid State Physics, University of Tokyo and the Information Processing Center, Saitama University. This work is supported by a Grant-in-Aid for Scientific Research from the Ministry of Education, Culture, Sports, Science and Technology, Japan. 

\appendix\section{}

In this appendix, we prove that our decimation procedure corresponds to the deflation process of the Fibonacci sequence.

\begin{figure}
\centerline{\includegraphics[width=80mm]{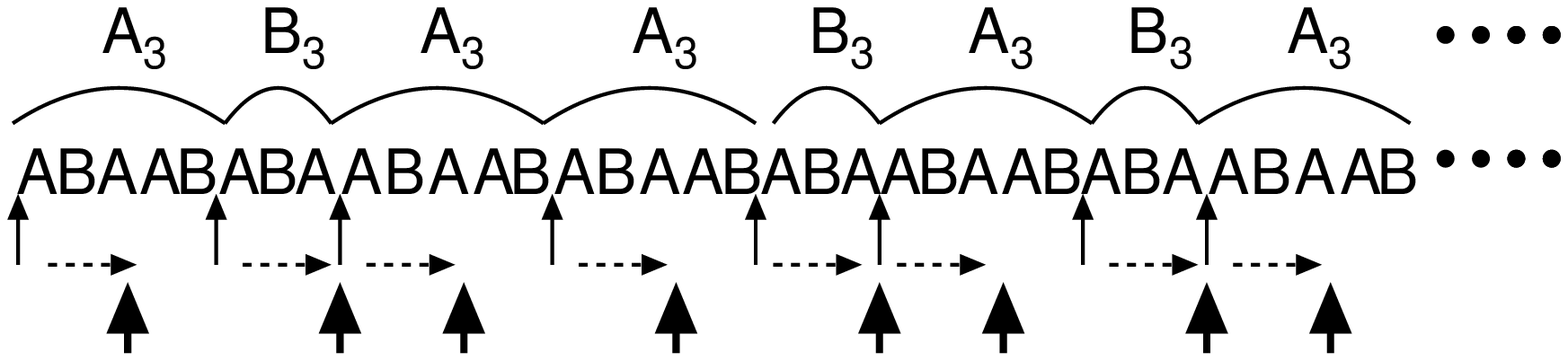}}
\caption{Construction of the Fibonacci sequence from $A_3$ and $B_3$. }
\label{appen}
\end{figure}

Let us build a Fibonacci sequence from the third generation building blocks A$_3\equiv$ ABAAB and B$_3 \equiv$ ABA as shown in Fig. \ref{appen}. The thin upward arrows indicate the boundary of these building blocks. Following the decimation process described in the text, the spins survive in the middle of the sequence AA as indicated by thick upward arrows. 

From the construction rule, the ABA sequences which are common to the left ends of the A$_3$ and B$_3$ arise at the right side of every block boundary. If the next block is A$_3$, the third position to the right of the boundary is always AA. If the next block is B$_3$, the next to next block is always A$_3$ because there appear no successive B's in the Fibonacci sequence. Therefore in this case also the third position to the right of the boundary is AA (next boundary). This implies that there always appear the sequence AA at the third position to the right of the boundary. Because these are the only cases in which AA sequence is allowed, at the third position to the {\it left} of the AA-sequence, we have always the block boundary. 

Thus, the sequence built by the substitution rule A $\rightarrow$ ABABA and B $\rightarrow$ ABA, which is the inverse of our decimation procedure, is a Fibonacci sequence lacking in the first three letters ABA. Therefore, if we apply our decimation procedure to the Fibonacci sequence, we have again a Fibonacci sequence with an additional B' at the leftmost position.


\begin{thebibliography}{11} 
\bibitem{shecht} D. Shechtman, I. Blech, D. Gratias and J. W. Cahn : Phys. Rev. Lett. {\bf 53} (1984) 1951.


\bibitem{kkt1} M. Kohmoto, L. P. Kadanoff and C. Tang : Phys. Rev. Lett {\bf 50} (1983) 1870.
\bibitem{ko1} M. Kohmoto and Y. Oono : Phys. Lett {\bf 102A} (1984) 145.

\bibitem{kst1} M. Kohmoto, B. Sutherland and C. Tang : Phys. Rev. B{\bf 35} (1987) 1020.

\bibitem{hk1} H. Hiramoto and M. Kohmoto : Int. J. Mod. Phys. B{\bf 6} (1992) 281 and references therein.
\bibitem{sato1} T. J. Sato, H. Takakura, A. P. Tsai and K. Shibata : Phys. Rev. Lett. {\bf 81} (1998) 2364.
\bibitem{sato2} T. J. Sato, H. Takakura, A. P. Tsai, K. Shibata, K. Ohoyama and K. H. Andersen : Phys. Rev. {\bf B61} (2000) 476.

\bibitem{jh1} J. Hermisson : J. Phys. A : Math. Gen. {\bf 33} (2000) 57.
\bibitem{kh1} K. Hida : J. Phys. Soc. Jpn. {\bf 68} (1999) 3177. 
\bibitem{kh2} K. Hida : J. Phys. Soc. Jpn. {\bf 69} (2000) Suppl. A 311.
\bibitem{kh3} K. Hida :   Phys. Rev. Lett. {\bf 93} (2004) 037205.
\bibitem{vidal1} J. Vidal, D. Mouhanna and T. Giamarchi : Phys. Rev. Lett. {\bf 83} (1999) 3908. 
\bibitem{vidal2} J. Vidal, D. Mouhanna and T. Giamarchi : Phys. Rev. {\bf B65} (2002) 014201.
\bibitem{vieira} A. P. Vieira : cond-mat/0403635.
\bibitem{acg} M. Arlego, D. C. Cabra, and M. D. Grynberg : Phys. Rev. {\bf B 64} (2001) 134419.
\bibitem{arl} M. Arlego : Phys. Rev. {\bf B 66} (2002) 052419. 
\bibitem{wjh} S. Wessel, A. Jagannathan and S. Haas: Phys. Rev. Lett. 90 (2003) 177205.
\bibitem{jag} A. Jagannathan, Phys. Rev. Lett. {\bf 92} (2004) 047202. 
\bibitem{df1} D. S. Fisher: Phys. Rev. B {\bf 50} (1994) 3799.
\bibitem{hen} C. L. Henley : Phys. Rev. Lett. {\bf 54} (1985) 2030.
\bibitem{her} C. K. Harris and R. B. Stinchcombe, Phys. Rev. Lett. {\bf 56} (1986) 869.

\bibitem{ram} R. Rammal and A. Benoit, Phys. Rev. Lett. {\bf 55} (1985) 649.
\bibitem{hy1} R. A. Hyman, K. Yang, R. N. Bhatt and S. M. Girvin: Phys. Rev.
Lett. {\bf 76} (1996) 839.
\bibitem{hy2} R. A. Hyman and K. Yang: Phys. Rev. Lett. {\bf 78} (1997) 1783.
\bibitem{wes} E. Westerberg, A. Furusaki, M. Sigrist, and P. A. Lee : Phys. Rev. Lett. {\bf 75} (1995) 4302. 
\end{thebibliography}
\end{document}